\documentclass[letterpaper, 10 pt, conference]{ieeeconf} 

\IEEEoverridecommandlockouts                
\usepackage{graphicx} % for pdf, bitmapped graphics files
\usepackage{mathptmx} % assumes new font selection scheme installed
\usepackage{times} % assumes new font selection scheme installed
\usepackage{amsmath} % assumes amsmath package installed
\usepackage{amssymb}  % assumes amsmath package installed
\usepackage{bm}
\usepackage{graphicx}
\usepackage{subcaption}
\usepackage{xcolor}

\title{\LARGE \bf
Hypersensitivity and Turnpikes in Optimal Control of Inverted Pendulum: A Dynamical Systems Perspective
}

\author{Mai Bando and Yuzuru Sato% <-this % stops a space
\thanks{
M.B. is supported by JSPS Grant-in-Aid for Scientific Research (B), JP No.23K24919 and JP No.26K02999.
Y.S. is supported by JSPS Grant-in-Aid for Scientific Research (B) JP No.21H01002 and JSPS Moonshot under Project No. JPMJMS2282-15. 
}% <-this % stops a space
\thanks{Mai Bando is with Department of Aeronautics and Astronautics, Kyushu University, 744 Motooka,
Nishu-ku, Fukuoka, 819-0395, Fukuoka, Japan
        {\tt\small mbando@aero.kyushu-u.ac.jp}}%
\thanks{Yuzuru Sato is with the RIES / Department of Mathematics, Hokkaido  University,
        N12 W7 Kita-ku, Sapporo, 060-0812 Hokkaido, Japan 
        {\tt\small ysato@math.sci.hokudai.ac.jp}}%
}

\begin{document}

\maketitle
\thispagestyle{empty}
\pagestyle{empty}

%%%%%%%%%%%%%%%%%%%%%%%%%%%%%%%%%%%%%%%%%%%%%%%%%%%%%%%%%%%%%%%%%%%%%%%%%%%%%%%%
\begin{abstract}
The hypersensitivity and turnpike phenomena in the optimal control of an inverted pendulum are investigated from a dynamical-systems perspective. We show that, for a fixed terminal time and a fixed terminal state optimal control problem, (1) the hypersensitivity originates from the fractal structure of the set of initial adjoint variables in the associated Hamiltonian dynamics, (2) the turnpike arises from slow dynamics in the vicinity of a degenerate center manifold, and (3) the escape channels are formed by normally hyperbolic invariant manifolds (NHIMs). As a consequence, small perturbations in the initial adjoint variables lead to qualitatively distinct extremal trajectories, resulting in severe numerical instability in trajectory optimization. Both the fractal structure and the invariant sets are characterized numerically and analytically.
\end{abstract}

%%%%%%%%%%%%%%%%%%%%%%%%%%%%%%%%%%%%%%%%%%%%%%%%%%%%%%%%%%%%%%%%%%%%%%%%%%%%%%%%
\section{Introduction}
The hypersensitivity of optimal control problems over long time intervals has been widely recognized \cite{anderson1987optimal,rao1999dichotomic}. In two-point boundary value problems (TPBVP), the terminal time and terminal state often exhibit exponential sensitivity with respect to perturbations in the initial adjoint variables, leading to severe numerical difficulties in trajectory optimization. While this phenomenon has been extensively studied from the viewpoints of numerical conditioning and singular perturbation theory \cite{kokotovic1999singular}, its intrinsic geometric origin in the underlying dynamical system remains not fully understood. Recent studies have suggested that such sensitivity is fundamentally linked to the Hamiltonian structure in optimal control problems ~\cite{hsiao2006fundamental,dell2020sensitivity}.

We show that hypersensitivity is closely related to the so-called turnpike phenomenon, in which optimal trajectories remain for most of the time near a specific region of the state space. In classical linear-quadratic problems, this region is typically a steady-state solution of an associated static optimization problem, and optimal trajectories converge exponentially toward this equilibrium under suitable controllability assumptions \cite{trelat2015turnpike}. However, recent developments have revealed that the turnpike is more naturally interpreted as a geometric object in the state space \cite{sakamoto2021turnpike}. In particular, it may correspond to a variety of invariant sets arising from the Hamiltonian structure in optimal control problems, rather than a single equilibrium point. 
In minimum-energy optimal control problems, where the cost does not penalize the state variables, the steady-state interpretation of the turnpike becomes degenerate. In such cases, the static optimization problem may fail to define a unique equilibrium, and the classical exponential turnpike property does not necessarily hold. 

In this paper, we revisit hypersensitivity and the turnpikes in the optimal control of the inverted pendulum from the perspective of dynamical systems theory. We show that hypersensitivity originates from a degenerate center acting as a turnpike, together with normally hyperbolic invariant manifolds (NHIMs) \cite{wiggins2013normally} that serve as escape channels. 
%For long escape times, 
These invariant sets and the fractal structure of possible initial states in the adjoint space to reach a neighborhood  of a given terminal state, which is also known as exit basins in Hamiltonian dynamics,  are analyzed both numerically and analytically.

\section{Hypersensitivity and turnpikes}
The hypersensitivity in optimal control has been widely studied in the context of numerical methods for trajectory optimization in dynamical systems \cite{bryson2018applied,betts1998survey,anderson1987optimal}. In TPBVP, the terminal state $q(t_f)$ with the terminal time $t_f$ 
may exhibit exponential sensitivity to the initial condition of the adjoint variable $\bm{p}(0)$ 
%\begin{equation}
%\frac{\partial q(t_f)}{\partial p(0)} \sim e^{t_f}, 
%\end{equation}
leading to severe numerical difficulties in trajectory optimization.
%This phenomenon can be interpreted in terms of a complex escape-basin structure in the underlying Hamiltonian dynamics.

 \begin{figure}[b]
      \centering
      \includegraphics[scale=0.1]{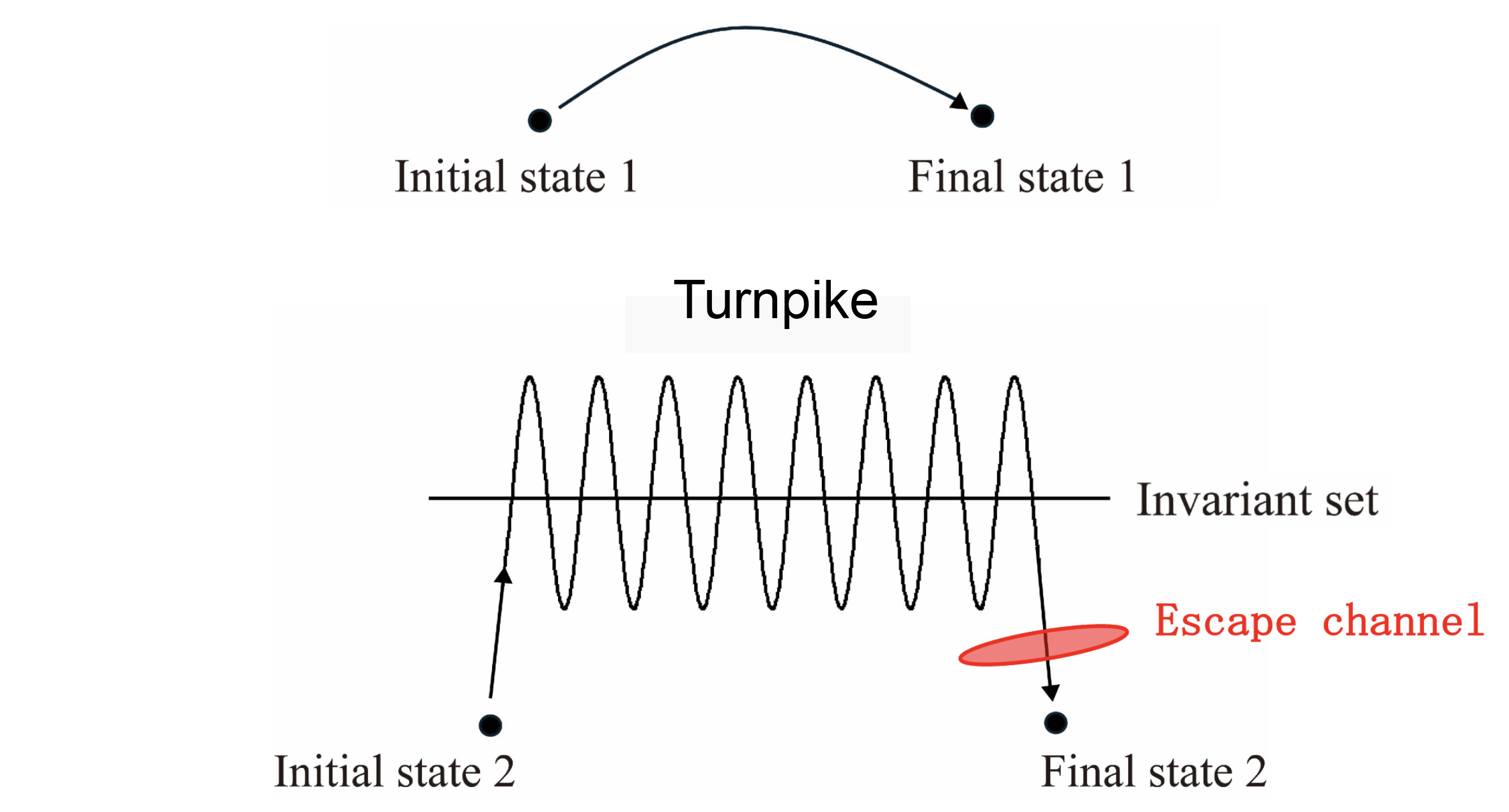}
      \caption{A schematic view of turnpike: A trajectory starting from an initial state 1 proceeds directly to  aterminal state 1, whereas a trajectory starting from an initial state 2 first approaches an invariant set and then escapes via escape channels to a terminal state 2.}
      \label{fig:turnpike}
\end{figure}

The turnpike is a geometric structure in state space that optimal trajectories typically approach and subsequently depart from.
When this structure corresponds to a saddle-type invariant set with unstable directions, trajectories can remain in its vicinity for extended periods before escaping (Fig. 1).

%\textcolor{red}{[Hypersensitivity and turnpikes should be explained more.]}

\section{Optimal control of the inverted pendulum}
\subsection{Minimum-energy problem of the inverted pendulum}

The dynamics of a pendulum subject to a control torque are given by
\begin{align}
\frac{d}{dt}
\begin{bmatrix}
q_1\\
q_2
\end{bmatrix}
=
\begin{bmatrix}
q_2\\
-k^2\sin q_1 + u
\end{bmatrix},
\quad
k=\sqrt{\frac{g}{l}},
\end{align}
where $\bm{q} = (q_1, q_2)^T$ denotes the costate vector, \(q_1\) is the angular displacement measured counterclockwise
from the downward vertical, \(q_2\) is the angular velocity, and \(u\) is the normalized control input. The state \((q_1,q_2)=(0,0)\) corresponds to the stable downward equilibrium and positive  
$q_1$, $q_2$ and $u$ correspond to the counterclockwise direction.

We consider the following fixed-terminal-time minimum-energy optimal control problem:
\begin{align}
\min_{u(\cdot)} \quad & J = \frac{1}{2} \int_{t_0}^{t_f} u^2(t)\, dt \\
\text{s.t.} \quad &
\dot{\bm{q}} = 
\begin{bmatrix}
q_2 \\ -k^2 \sin q_1 + u
\end{bmatrix}, \\
& \bm{q}(t_0) = \bm{q}_0,\quad \bm{q}(t_f) = \bm{q}_f.
\end{align}
According to Pontryagin’s maximum principle (PMP) \cite{pontryagin1962mathematical,bryson2018applied}, there exists an adjoint vector $\bm{p}(t)$ such that the extremals satisfy the canonical Hamiltonian system together with the stationarity condition with respect to the control input.
The corresponding pre-Hamiltonian is defined as
\begin{equation}
\bar{H}(\bm{q},\bm{p},u) = \frac{1}{2} u^2 + p_1 q_2 + p_2\left(-k^2\sin{q_1} + u \right),
\end{equation}
where $\bm{p} = (p_1, p_2)^T$ denotes the costate vector. The stationarity condition $\partial \bar{H}/\partial u = 0$ yields the minimizing control
\begin{align}
u^* = -p_2.
\end{align}
Substituting this expression into the pre-Hamiltonian leads to the reduced Hamiltonian
\begin{align}
H(\bm{q},\bm{p}) = -\frac{1}{2} p_2^2 + p_1 q_2 - k^2 p_2 \sin{q_1}.
\end{align}
The canonical equations governing the extremals are then given by
\begin{align}
 &\dot{\bm{q}}=\left(\frac{\partial H}{\partial \bm{p}}\right)^T=    \begin{bmatrix}
        q_2 \\-k^2\sin{q_1} -p_2
    \end{bmatrix}\\
 &\dot{\bm{p}}=-\left(\frac{\partial H}{\partial \bm{q}}\right)^T
=    \begin{bmatrix}
        k^2 p_2 \cos q_1\\ -p_1
    \end{bmatrix}
\end{align}
These equations define a four-dimensional Hamiltonian system describing the extremals of the minimum-energy control problem.
This yields a two-point boundary value problem (TPBVP) in the state–adjoint variables, where the boundary conditions are imposed only on the state variables:
\begin{equation}
\bm{q}(t_0)=\bm{q}_0,\quad \bm{q}(t_f)=\bm{q}_f,
\end{equation}
while the initial values of the adjoint variables must be determined so as to satisfy the terminal constraints \cite{keller1976numerical}.

\subsection{Equilibria, Stability, and Periodic Orbit Families}
For $k=1$, there are four types of equilibria: $(q_1,q_2,p_1,p_2) =(0,0,0,0)$, $(\pm\pi,0,0,0)$, and the two symmetric equilibria $(\pi/2,0,0,-1)$ and $(-\pi/2,0,0,1)$. 
The origin is a degenerate center with Jacobian eigenvalues $\lambda=\pm i$ (double roots), each with algebraic multiplicity two. The equilibria $(\pm\pi,0,0,0)$ are non-semisimple hyperbolic saddles with eigenvalues $\lambda=\pm1$ (double roots), each with algebraic multiplicity two. The remaining two equilibria are of saddle$\times$center type, with eigenvalues $\lambda=\pm1,~\pm i$.

\subsection{Degenerate center at the origin and spiral-like slow escape}

%\textcolor{red}{Spiral}
At the origin $(q_1,q_2,p_1,p_2) = (0,0,0,0)$, the linearized system possesses purely imaginary eigenvalues with multiplicity, resulting in a degenerate center. As a consequence, trajectories starting near $q_1 = 0$ exhibit a spiral-like slow escape behavior, gradually drifting linearly with time away from the equilibrium . In particular, the region near $q_1 = 0$, trajectories remain for a relatively long time before escaping along weakly unstable directions. 

\subsection{Transit structure near the saddle$\times$center equilibrium}
\begin{figure}[h]
  \centering
  \begin{subfigure}[t]{0.31\textwidth}
    \centering
    \includegraphics[width=\textwidth]{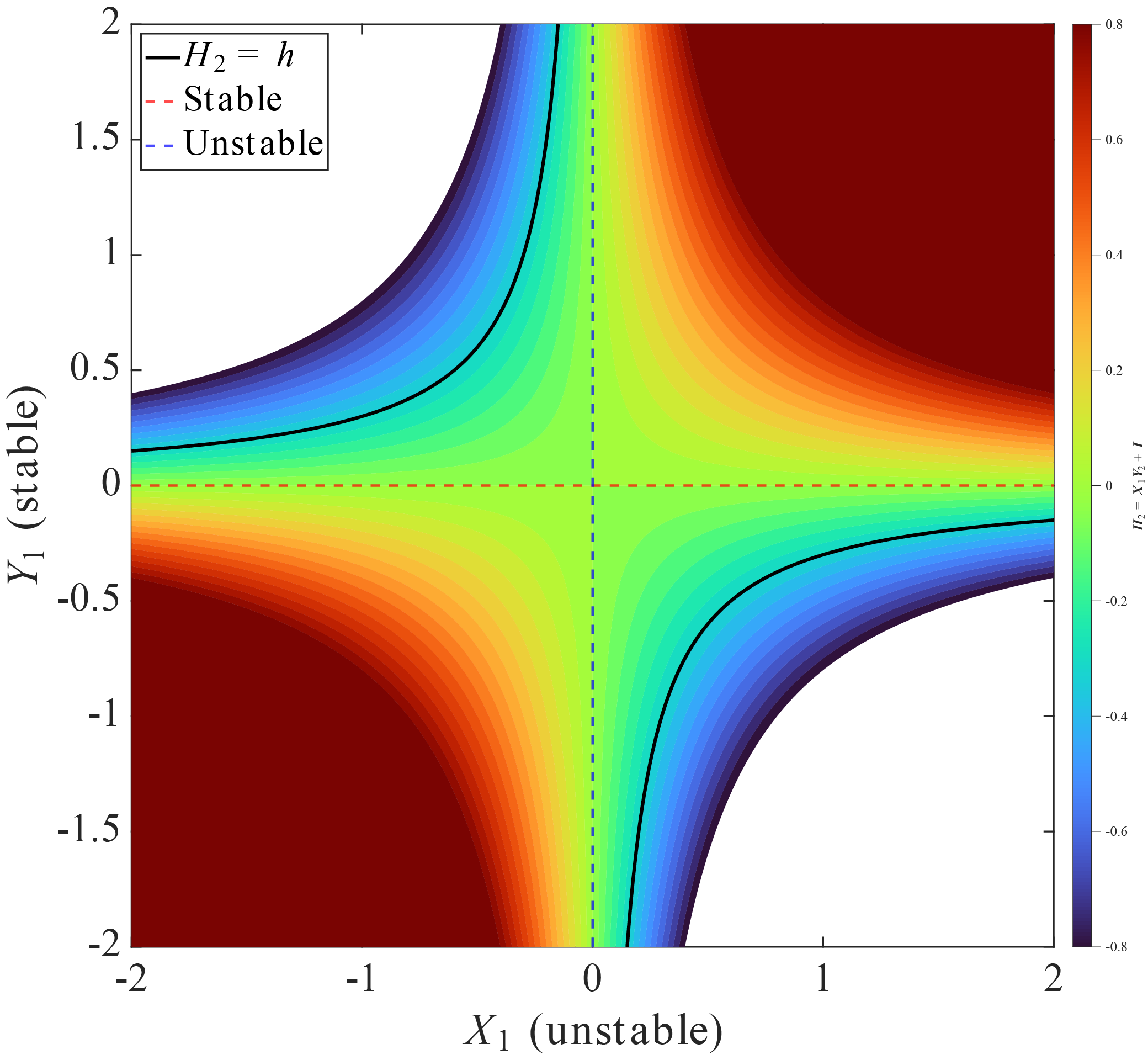}
    \caption{$I<-h$: non--transit}
  \end{subfigure}\hfill
  \begin{subfigure}[t]{0.31\textwidth}
    \centering
     \includegraphics[width=\textwidth]{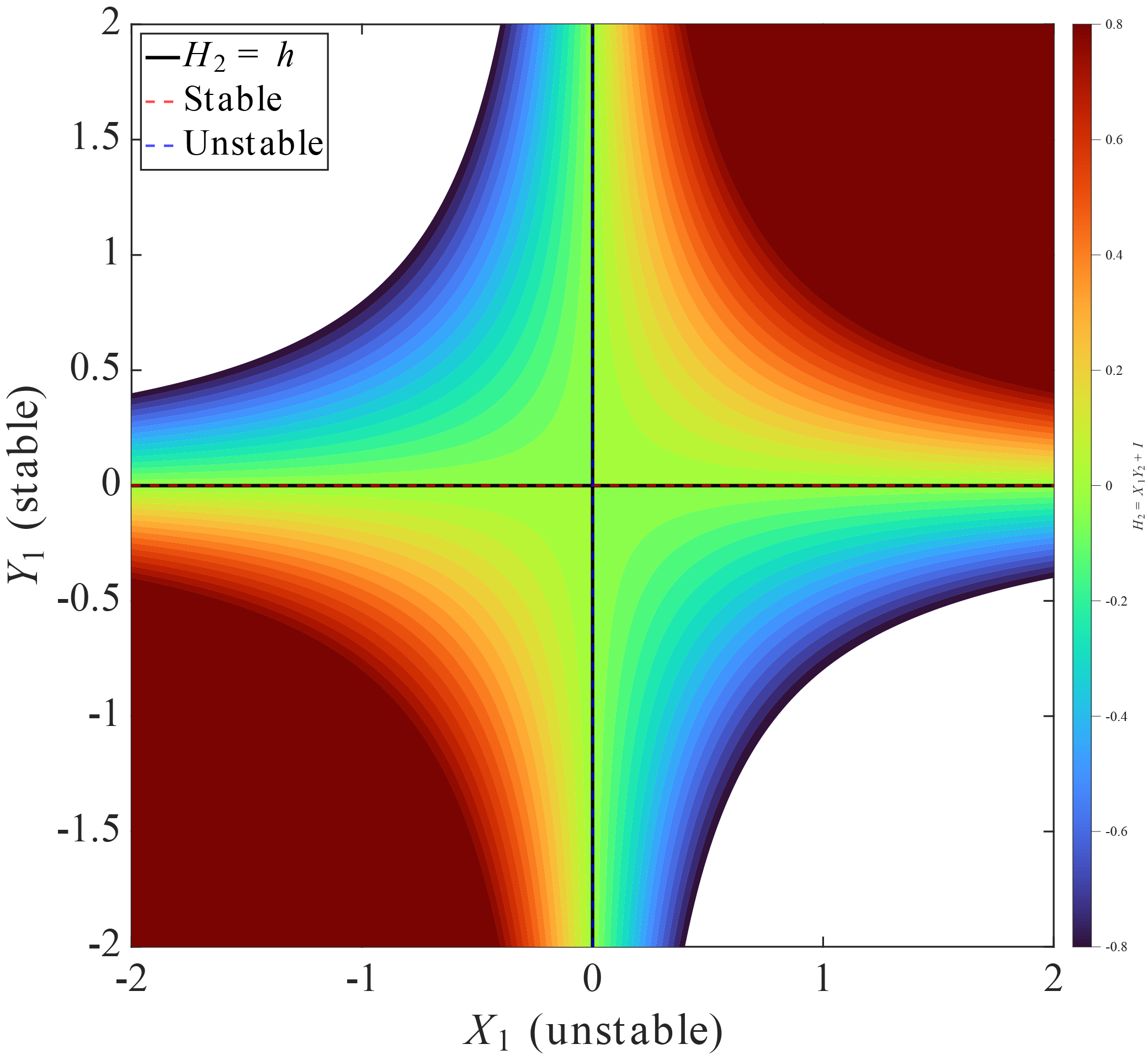}
    \caption{$I=-h$: asymptotic boundary}
  \end{subfigure}\hfill
  \begin{subfigure}[t]{0.31\textwidth}
    \centering
    \includegraphics[width=\textwidth]{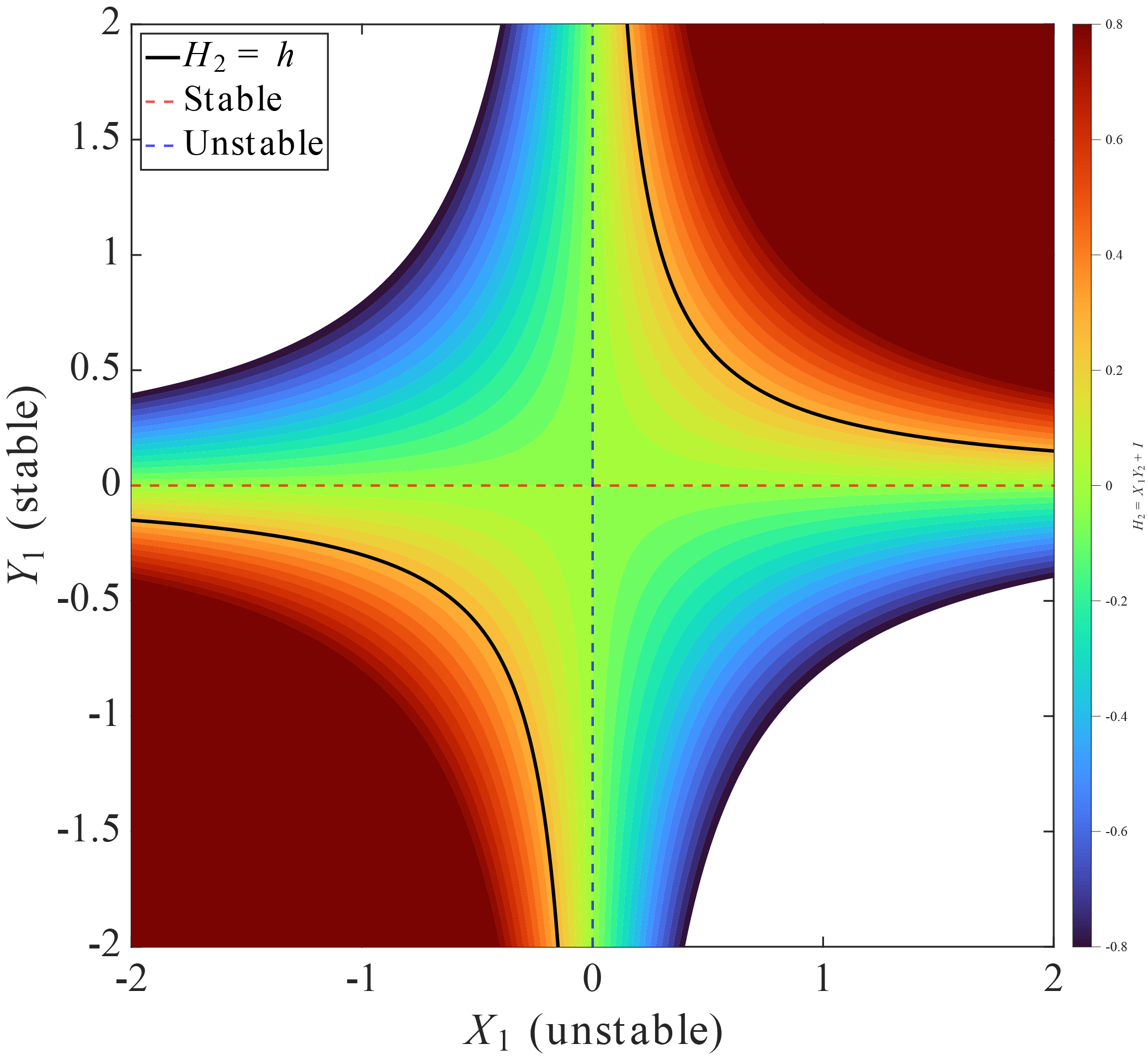}
    \caption{$I>-h$: transit}
  \end{subfigure}
  \caption{
  Local transit geometry on a fixed energy surface $H_2=h<0$.
  }
  \label{fig:transit-geometry}
\end{figure}

Around the equilibrium $x_C^*=\left(\pm \frac{\pi}{2},0,0,\mp 1\right)$,
the quadratic part of the Hamiltonian can be reduced, by a linear canonical transformation, to the saddle$\times$center normal form
\begin{equation}
H_2 = X_1 Y_1 - \frac{1}{2}\left(X_2^2 + Y_2^2\right),
\end{equation}
where $(X_1,Y_1)$ represent the hyperbolic unstable/stable directions and $(X_2,Y_2)$ represent the center mode.
Defining the center action
\begin{equation}
I := \frac{1}{2}\left(X_2^2 + Y_2^2\right)\ge 0,
\end{equation}
the fixed-energy condition $H_2=h$ becomes $X_1Y_1 = h + I$.
This relation determines the local transit structure in phase space \cite{conley1968low}. In particular,
\begin{equation}
\begin{cases}
h+I>0 & \text{: transit orbit},\\
h+I=0 & \text{: asymptotic boundary},\\
h+I<0 & \text{: non-transit orbit}.
\end{cases}
\end{equation}
The invariant manifolds $X_1=0,~Y_1=0$ form the local separatrix. For $h<0$, the NHIM is given by $X_1=Y_1=0,~I=-h$. 
Thus, the sign of $h$ alone does not determine whether a trajectory is transit or non-transit; the center action $I$ must also be taken into account.
Figure~\ref{fig:transit-geometry} illustrates the local transit geometry in the hyperbolic subspace $(X_1,Y_1)$.

%%%%%%%%%%%%%%%%%%%%%%%%%%%%%%%%%%%%%%%%%%%%%%%%%%%%%%%%%%%%%%%%%%%%%%%%%%%%%%%%

\section{Typical trajectories in optimal control of inverted pendulum}

\begin{figure}[thpb]
     \centering
\includegraphics[scale=0.25]{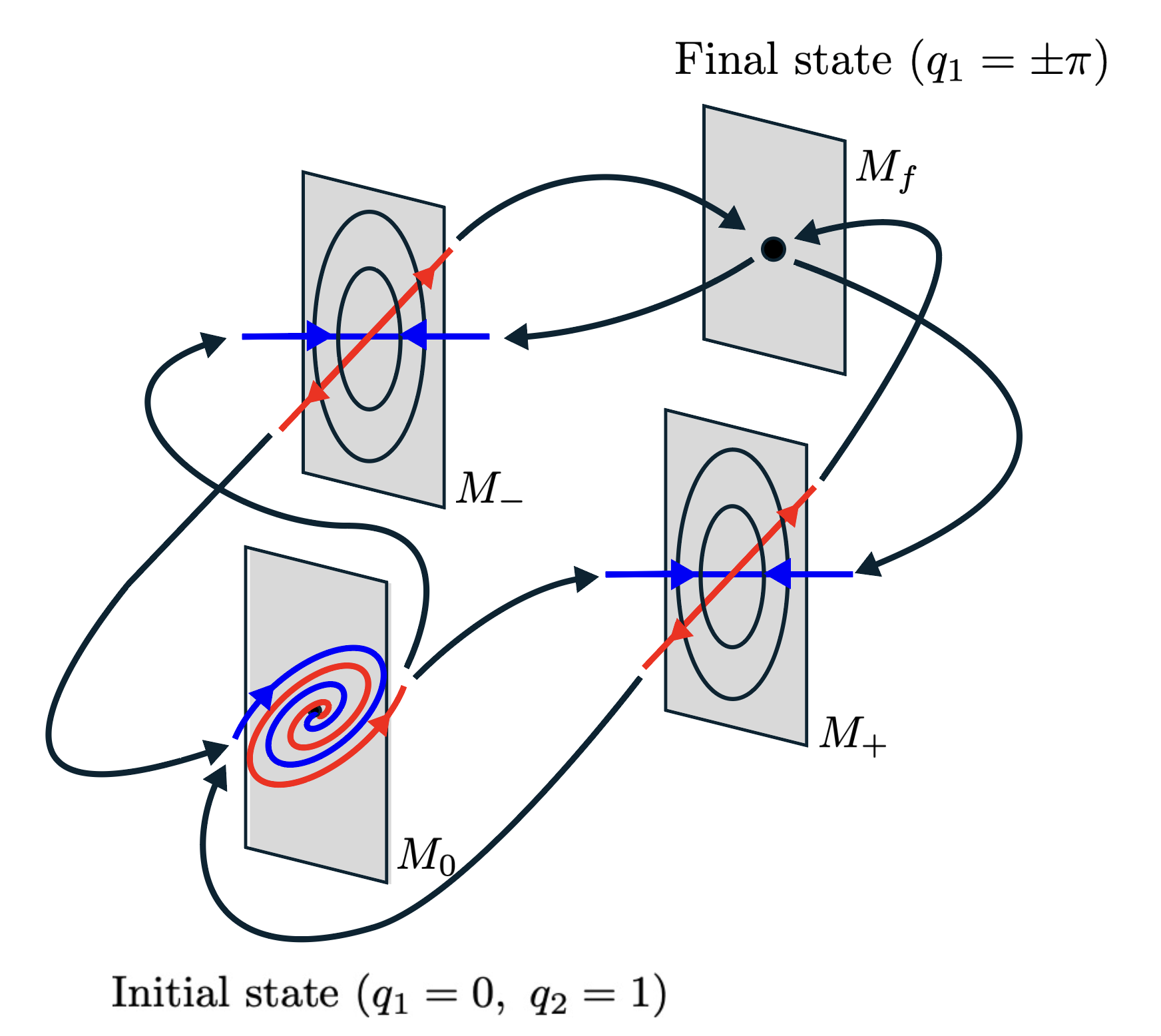}
     \caption{A schematic view of the global network of state space objects}
    \label{fig:snhims}
\end{figure}

We consider a fixed-terminal-time fixed-terminal-state optimal control problem in which the initial and terminal states are specified as $(q_1(0), q_2(0))=(0.0, 1.0)$ and  $(q_1(t_f), q_2(t_f))=(\pm\pi, q_2^*)$, respectively, where the terminal time $t_f$ and the terminal state $q_2^*$ are treated as free control parameters. 
%We assume that the system starts with an initial state $(q_1(0), q_2(0))=(0.0, 1.0)$ and reaches the final state $(q_1(t_f), q_2(t_f))=(\pm\pi, q_2^*)$. %, where $\epsilon=10^{-1}$. 
%The adjoint variables 
%$p_1, ~p_2$ are not restricted, and 
%The terminal time $t_f$ is a control parameter for the optimal trajectories. 
%We consider all problems corresponding to  $q_2^*\in{\bf R}$, and analyse a set of all possible control histories from . 
In this setting, 
%for  typical trajectories to reach the final state $q(t_f)$ at time $t_f$, 
the degenerate center works as turnpikes to generate long sustainable transients, and NHIMs work as escape channels to the terminal state. A global view of the network of NHIMs is schematically shown in Fig. \ref{fig:snhims}. In Fig. \ref{fig:snhims}, 
the set $M_0$ is defined as a cross section of a subspace containing the degenerate center $(0,0,0,0)$, with the section specified by the initial states  $(0,1,*,*)$. The stable and unstable spiral-like slow escapes starting from all admissible initial points on $M_0$ are schematically depicted by blue and red lines, respectively. The subspace $M_{\pm}$ includes the center$\times$saddle equilibrium point $(\pm\frac{\pi}{2},0,0,\mp 1)$ and associated NHIMs. The stable and unstable tubes starting from the NHIMs are schematically depicted by blue and red lines, respectively. The subspace $M_f$ includes the saddle$\times$saddle equilibrium point $(\pm\pi,0,0,0)$ and all admissible terminal states. The black lines with arrows represent possible routes to each subspaces.

At the initial stage, trajectories starting inside elliptic islands remain trapped and do not escape from the integrable torus. Since such islands may be distributed near the point $(0,1,*,*)$, initial adjoint variables must avoid them in order to reach the terminal state. For large terminal times $t_f$, trajectories are required to approach to the turnpikes, giving rise to long transients. They remain close to the turnpike region for extended periods before eventually escaping through specific channels, associated with the NHIMs $M_{\pm}$. %located near $q_1=\pm\frac{\pi}{2}$. 
After crossing $M_{\pm}$, the trajectories proceed toward the the terminal states $q_1=\pm\pi$ (Fig.~\ref{fig:nhims}).

The sojourn time near the turnpikes is extremely sensitive to the initial conditions, as trajectories must remain in this region for a precisely tuned duration. Once they cross the separatrices (i.e., the NHIMs), they rapidly transit to the terminal state.

%\begin{figure}[thpb]
%      \centering
%\includegraphics[scale=0.1]{figs/p_mani_149.jpg}\\
%\includegraphics[scale=0.1]{figs/q_mani_149.jpg}
%      \caption{NHIM}
%      \label{fig:nhim}
%\end{figure}

\begin{figure}[htbp]
    % \centering
    % \begin{subfigure}{0.9\linewidth}
    %     \centering
    %     \includegraphics[width=0.6\linewidth]{figs/spiral2d.jpg}
    %     \caption{Spiral-like slow escape behavior in the phase space projected onto the plane $(q_1, q_2)$.}
    %     \label{fig:spiral2d}
    % \end{subfigure}
    % \vspace{2mm}
%    \begin{subfigure}{0.9\linewidth}
        \centering
        \includegraphics[width=0.95\linewidth]{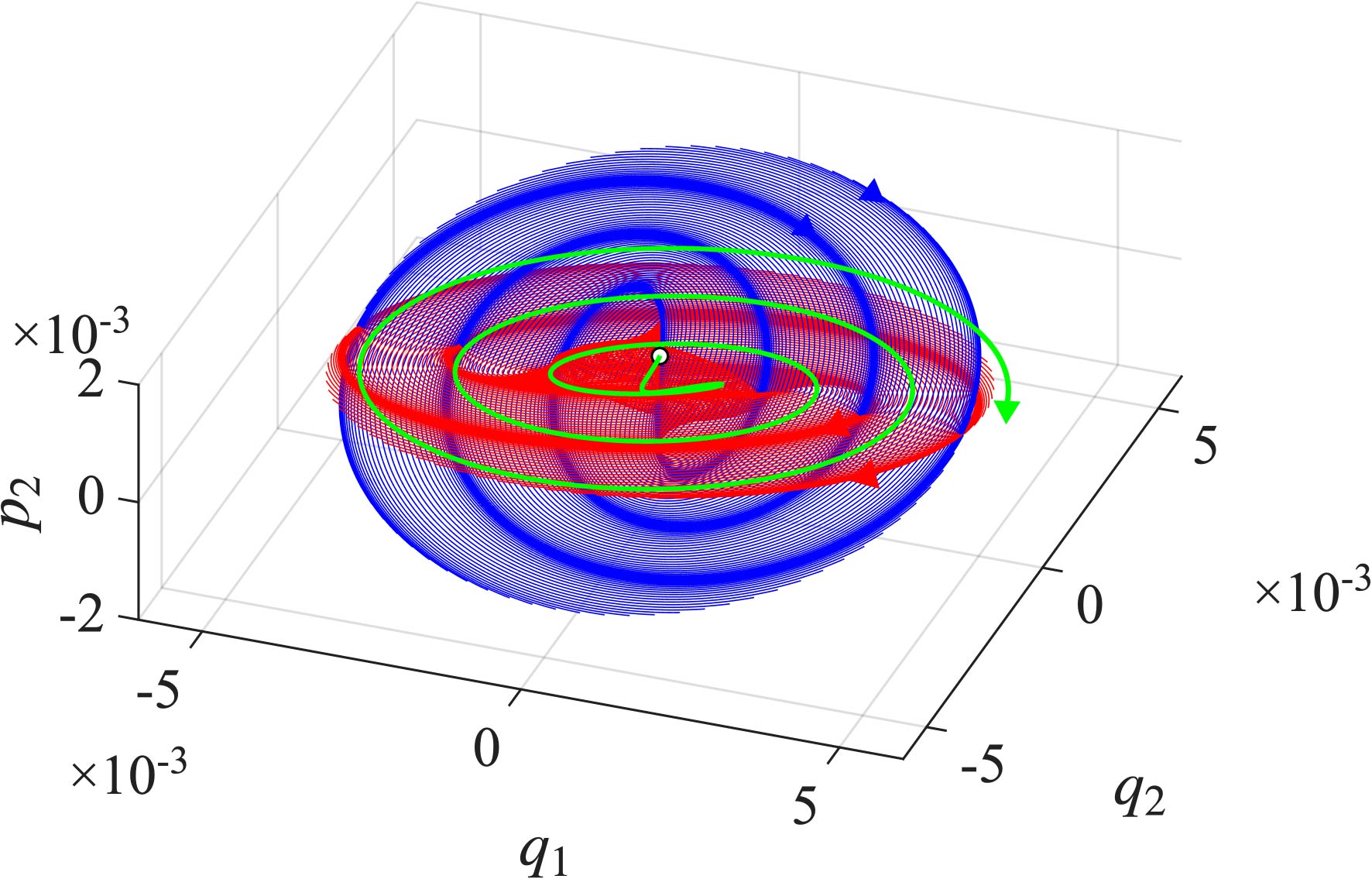}
 %       \caption{Spiral-like slow escape behavior in the phase space projected onto the subspace $(q_1, q_2, p_2)$.}
 %       \label{fig:spiral_full}
%    \end{subfigure}
        \caption{Spiral-like slow escape: The Jacobians at all admissible initial states $q_1=0.0$ have purely imaginary eigenvalues of multiplicity two, generating spiral-like slow escapes. The blue and red curves denote stable and unstable orbits, respectively, originating from the initial states within an $\epsilon$-ball centered at the origin, where $\epsilon=0.01$. }
    \label{fig:spiral_subfig}
\end{figure}

\begin{figure}[t]
    \centering
    % \begin{subfigure}{0.9\linewidth}
    %     \centering
    %     \includegraphics[width=0.7\linewidth]{figs/q_mani2d_new.jpg}
    %     \caption{Transit orbit in the phase space projected onto the plane $(q_1, q_2)$.}
    %     \label{fig:transit2d}
    % \end{subfigure}
    % \vspace{2mm}
%    \begin{subfigure}{0.95\linewidth}
        \centering
        \includegraphics[width=0.8\linewidth]{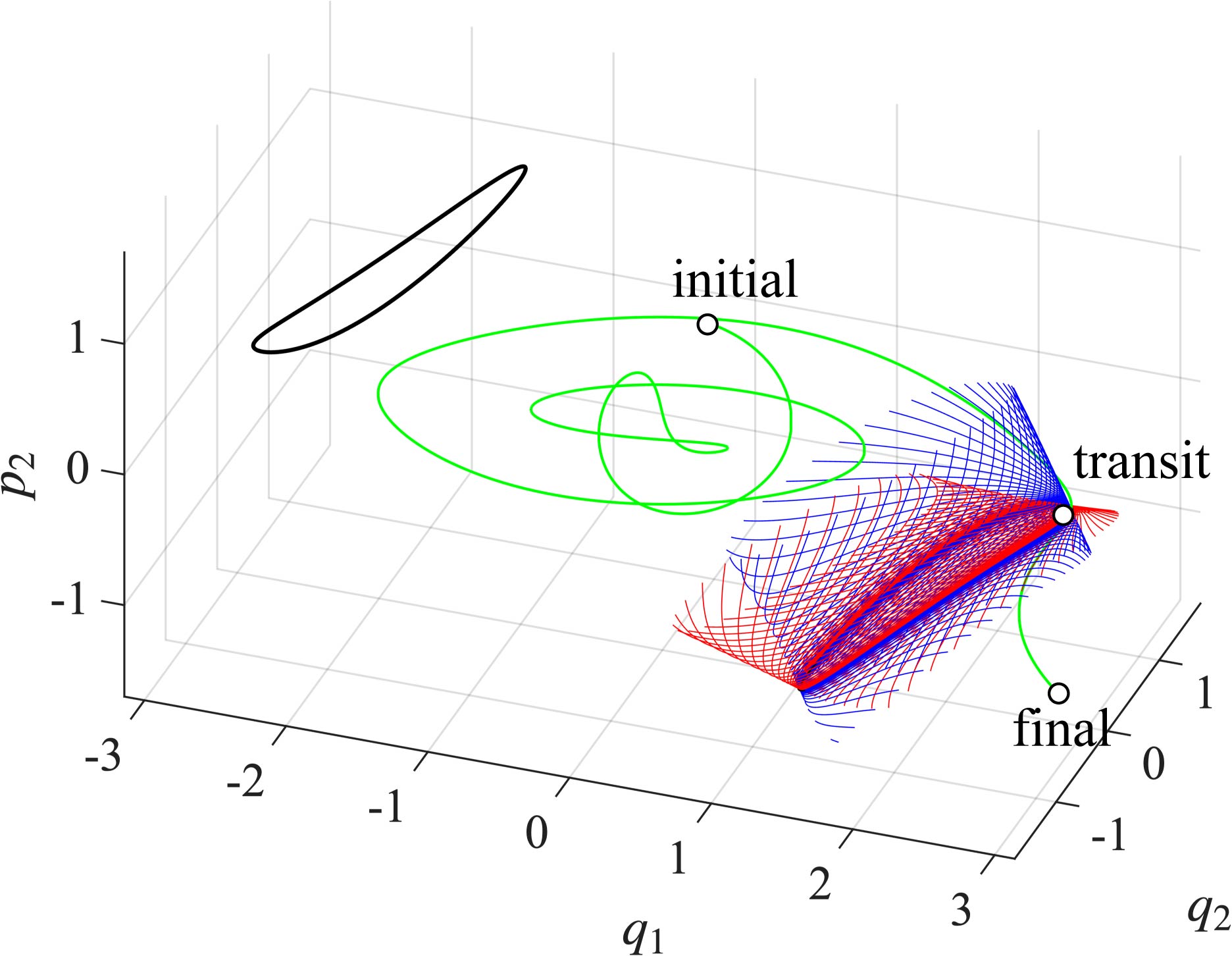}
 %       \caption{Transit orbit in the phase space projected onto the subspace $(q_1, q_2, p_2)$.}
 %       \label{fig:transit3d}
%    \end{subfigure}
              \caption{Transit orbit: An example of the trajectory started from $(q_1, q_2, p_1, p_2)=(0, 1.0, 0.05028, 0.2450)$ with a Hamiltonian $H=0.02025$ and $t_f=31.22$ is depicted in a green line with the initial, transit, and terminal states in white circles. NHIMs associated with the center at $q_1=\pm\frac{\pi}{2}$ are shown as periodic orbits in black lines, and the tubes, a bundle of stable / unstable manifold starting from the NHIM, are depicted in blue / red lines.}
    \label{fig:nhims}
\end{figure}

As an illustrative example, the degenerate center, the NHIMs, and their neighborhoods in the phase space of the Hamiltonian system with $H=0.02025$ and $t_f=31.22$ are shown in Figs.~\ref{fig:spiral_subfig} and \ref{fig:nhims}. The local saddle\(\times\)center analysis predicts a family of periodic-orbit NHIMs emanating from the equilibria at \(q_1=\pm \pi/2\) for energy levels \(H<1/2\) sufficiently close to \(H=1/2\). Numerical continuation of the full nonlinear Hamiltonian system shows that these families persist toward lower energy levels and can be continued toward \(H\to0^+\).

\begin{figure}[t]
    \centering
    \begin{subfigure}{0.9\linewidth}
        \centering
          \includegraphics[scale=0.38]{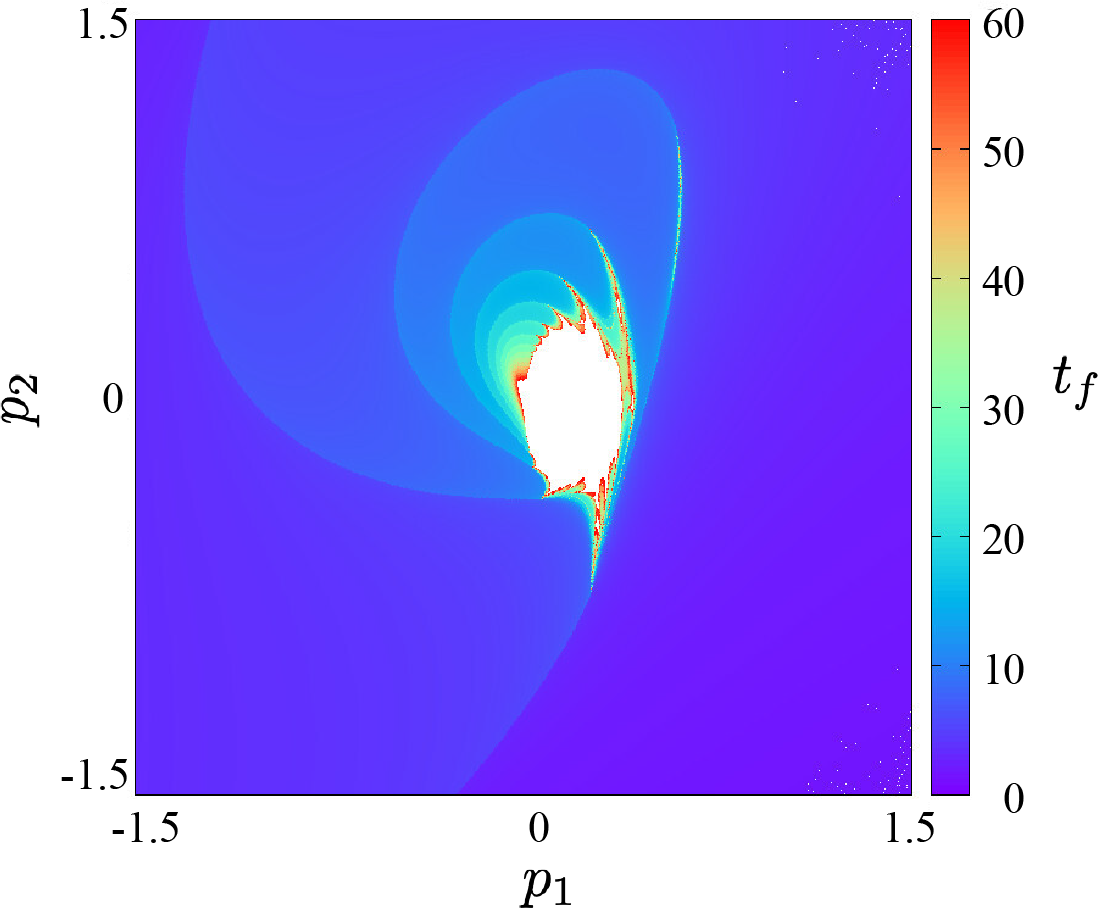}
        \caption{Terminal  time diagram for $(p_1(0), p_2(0))$}\vspace{2mm}
        \label{fig:escapetfa}
    \end{subfigure}
    \begin{subfigure}{0.9\linewidth}
    \centering
    \includegraphics[scale=0.38]{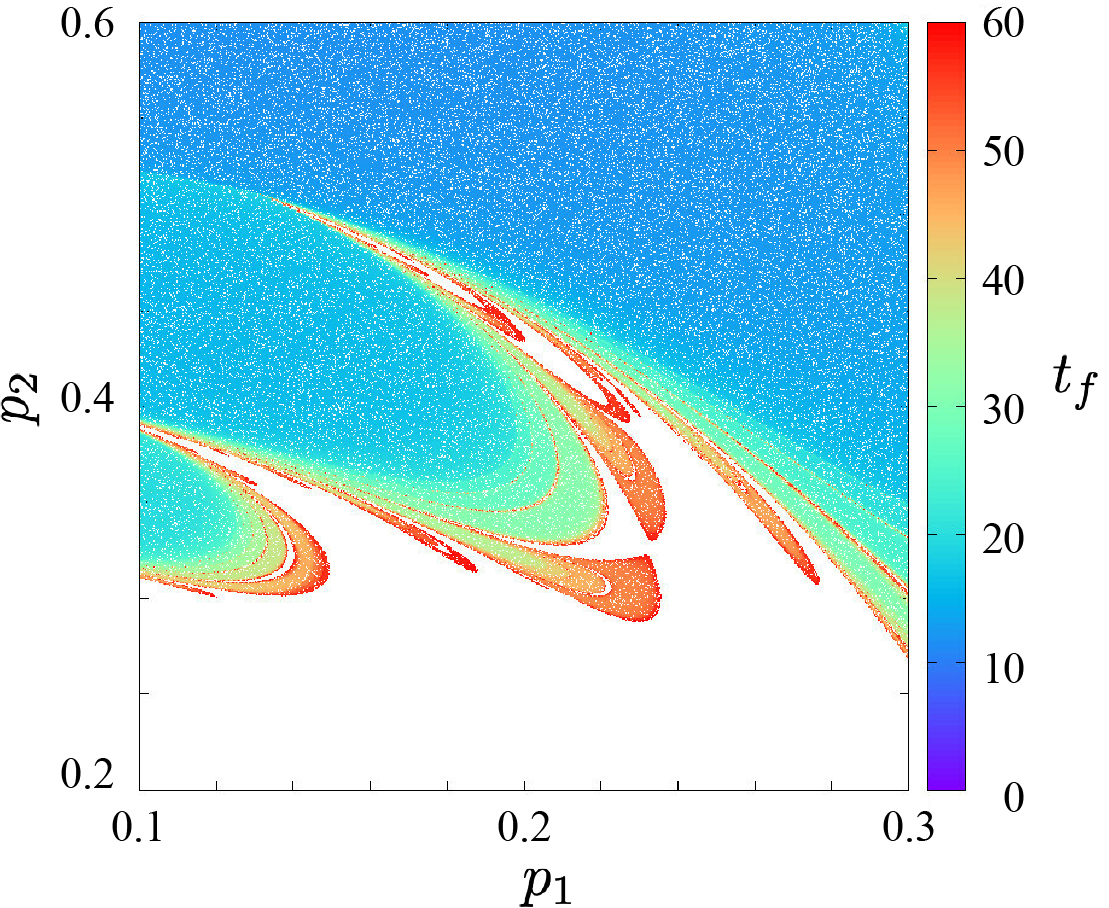}
      \caption{Enlargement of (a).} 
      \label{fig:escapetfb}
    \end{subfigure}
      \caption{Terminal time diagram and  basins: (a) Terminal time $t_f$, obtained from $1.0\times 10^{10}$ initial states $(0.0, 1.0, p_1,p_2)$, are represented by colors. The white regions represent the initial states with $t_f>60$ including integrable tori with $t_f\to\infty$.  
      %(b) The contour plot of the initial condition with the terminal time $t_f=10,20,30,40,50,60 \pm \delta t$, where $\delta t=0.1$, are plotted by colored points. The set of the same terminal time forms terminal time  basins.
      (b) Enlargement of (a): Fractal property is observed in the red region $t_f\in[50,60]$.} 
    \label{fig:escapetf}
\end{figure}

Here we consider a family of terminal conditions $(q_1(t_f),q_2(t_f))=(\pi,q_2^*)$, parameterized by $q_2^* \in \mathbb{R}$.
To examine their dependence on the initial adjoint variables, trajectories are propagated from the fixed initial state $(q_1(0),q_2(0))=(0,1)$ until they reach the terminal section $q_1=\pi$. The corresponding terminal time $t_f$ and terminal state $q_2(t_f)$ are then represented in the $(p_1(0),p_2(0))$ plane. 

\begin{figure}[thpb]
    \centering
    \begin{subfigure}{0.9\linewidth}
        \centering
        \includegraphics[scale=0.38]{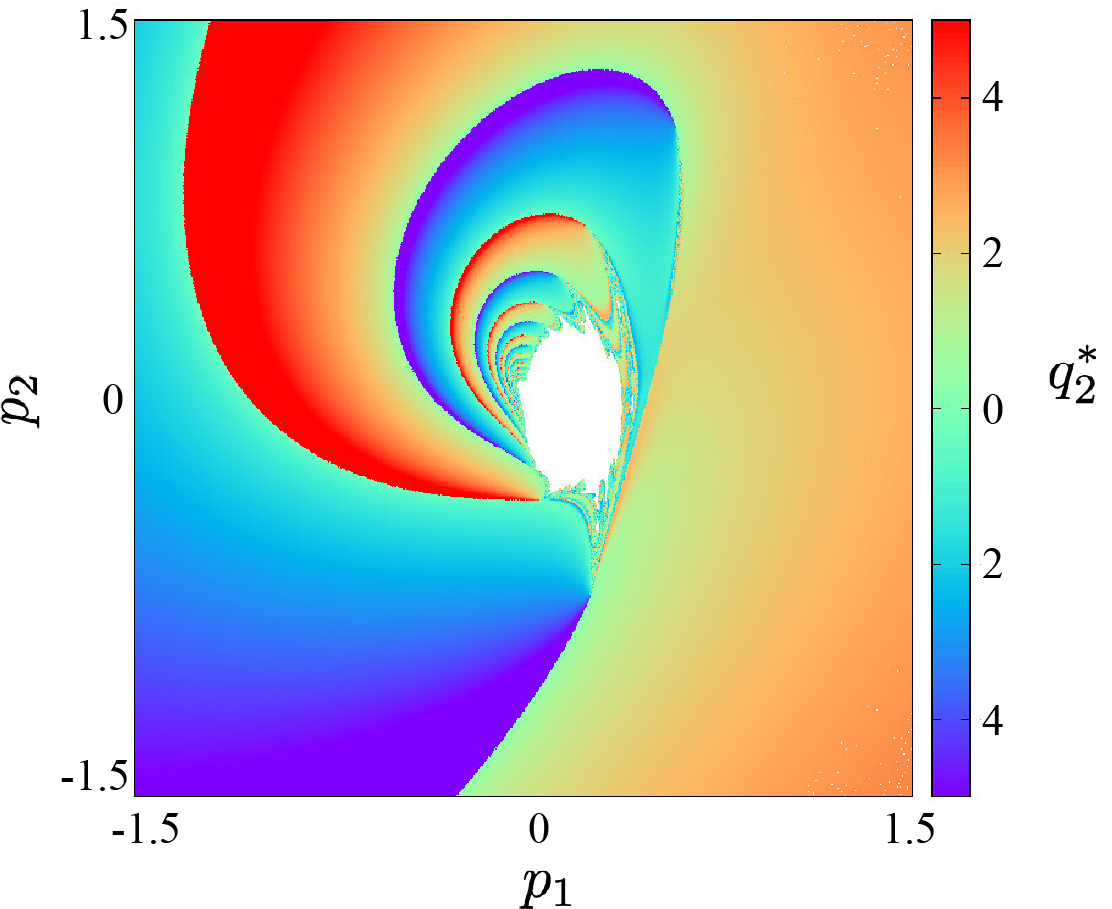}
        \caption{Terminal state diagram for $(p_1(0), p_2(0))$}\vspace{2mm}
        \label{fig:escapeq2a}
    \end{subfigure}
    \begin{subfigure}{0.9\linewidth}
    \centering
    \includegraphics[scale=0.38]{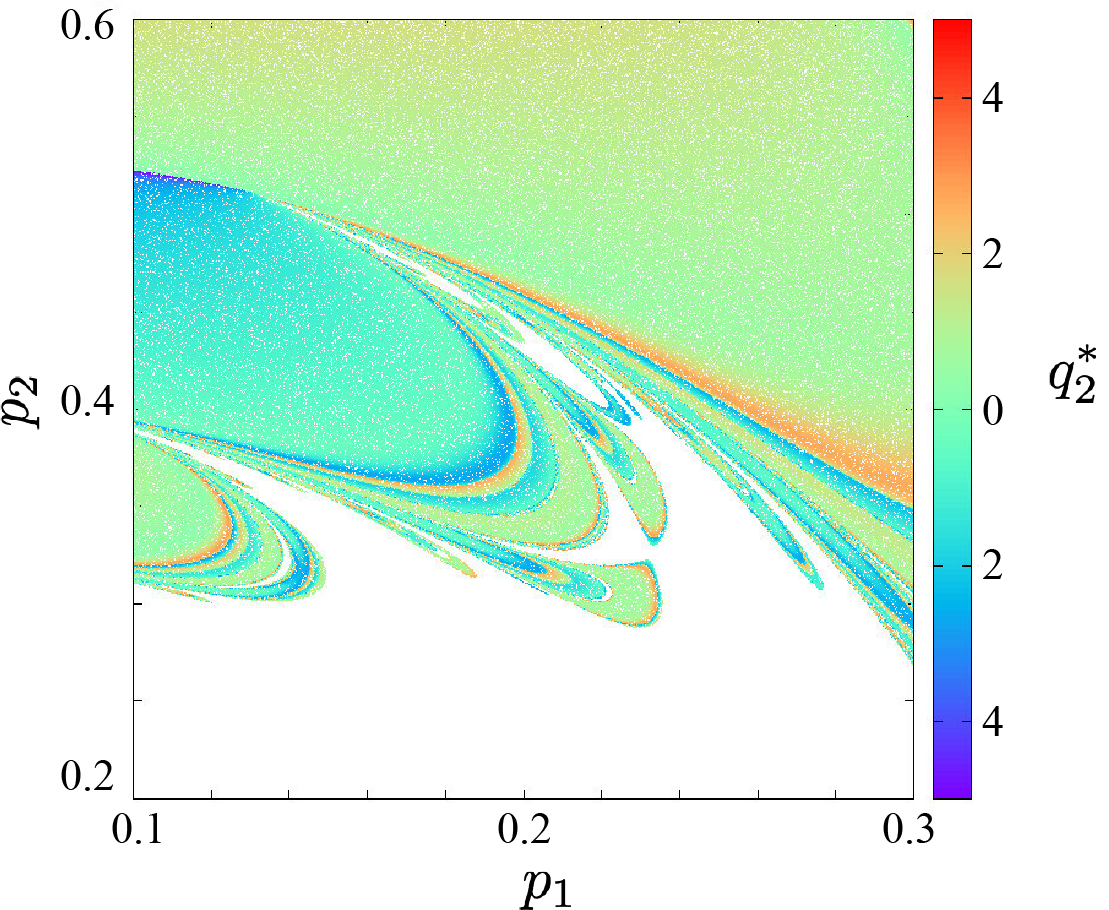}
      \caption{Enlargement of (a).} 
      \label{fig:escapeq2b}
    \end{subfigure}
      \caption{Terminal state diagram and  basins: (a) Terminal state  $q_2(t_f)$, obtained from $1.0\times 10^{10}$ initial states $(0.0, 1.0, p_1,p_2)$, are represented by colors. The white regions represent the initial states with $t_f>60$ including integrable tori with $t_f\to\infty$.  (b) Enlargement of (a).} 
    \label{fig:escapeq2}
\end{figure}

Figures~\ref{fig:escapetf} and \ref{fig:escapeq2} show the terminal time and terminal state diagrams, respectively, in the $(p_1(0),p_2(0))$ plane. The terminal time $t_f$ and terminal velocity $q_2^*$ are encoded by color. 
The white regions in the diagram correspond to initial states with terminal time $t_f>60$, including integrable tori for which trajectories never reach the terminal state, i.e., $t_f\to\infty$. 
Both diagrams exhibit intricate, fractal-like structures, indicating a strong sensitivity of the terminal conditions to the initial adjoint variables. In particular, small perturbations near these structures can produce large variations in both $t_f$ and $q_2^*$.
The enlarged views further reveal the fine structure of these regions, analogous to exit-basin boundaries in open Hamiltonian systems \cite{aguirre2009fractal}.

\begin{figure}[thbp]
    \centering
    \begin{subfigure}{0.9\linewidth}
        \centering
        \includegraphics[scale=0.38]{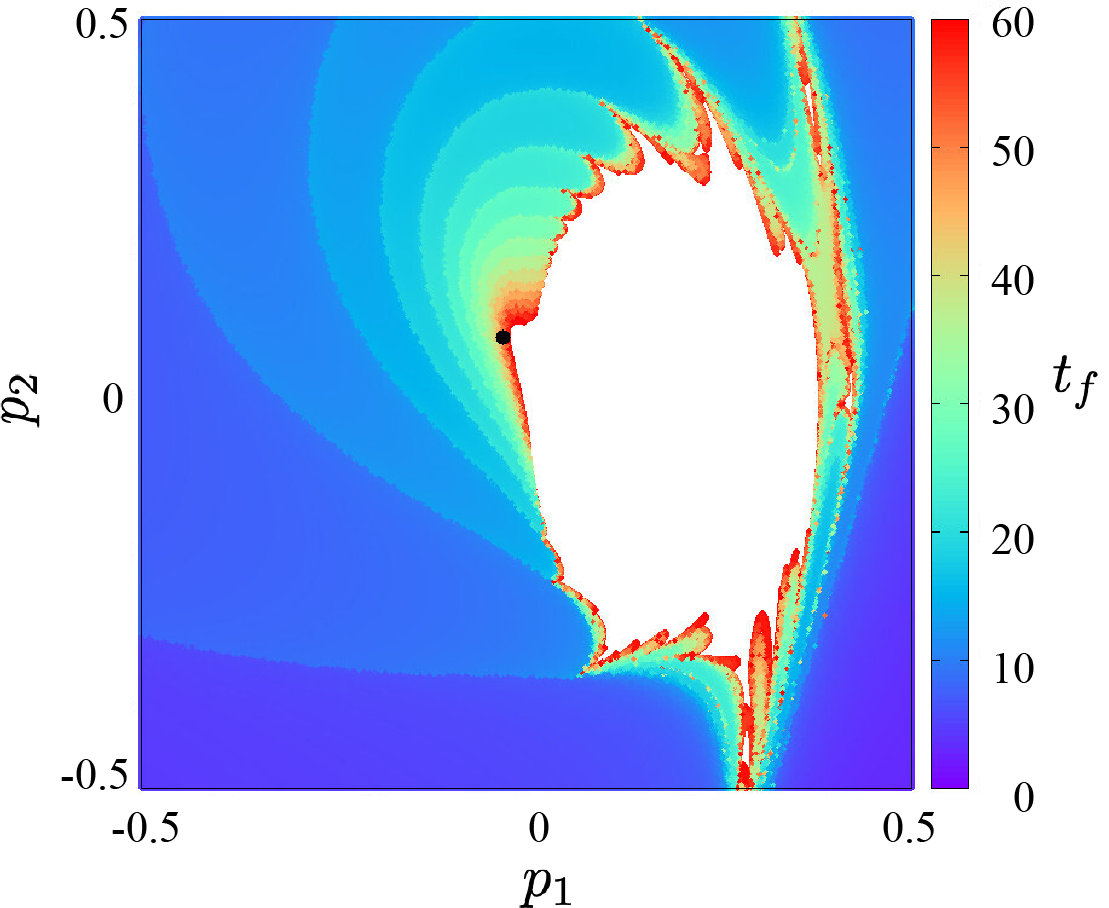}
        \caption{Enlargement of the terminal-time diagram.}\vspace{2mm}
        \label{fig:tpbvp1}
    \end{subfigure}
    \begin{subfigure}{0.9\linewidth}
    \centering
    \includegraphics[scale=0.38]{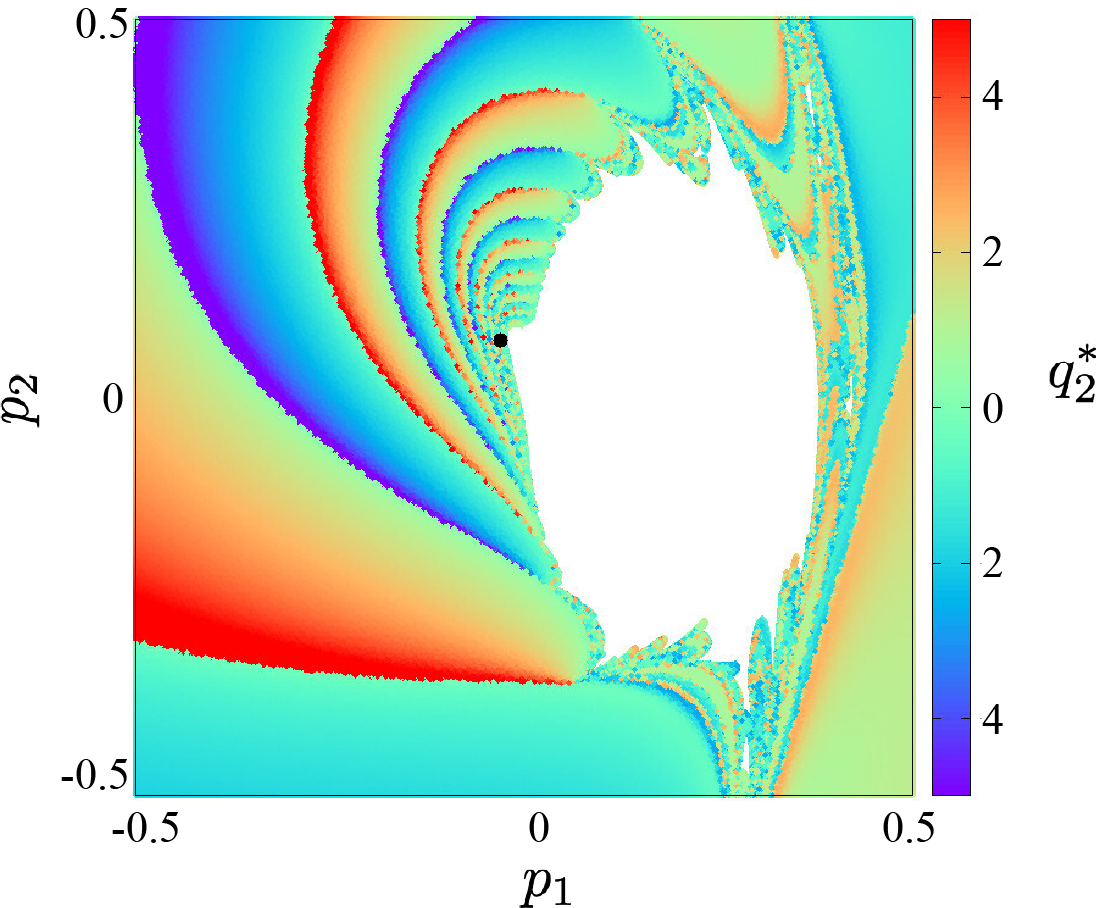}
      \caption{Enlargement of the terminal-state diagram.} 
      \vspace{2mm}
      \label{fig:tpbvp2}
    \end{subfigure}
        \begin{subfigure}{0.9\linewidth}
    \centering
    \includegraphics[scale=0.38]{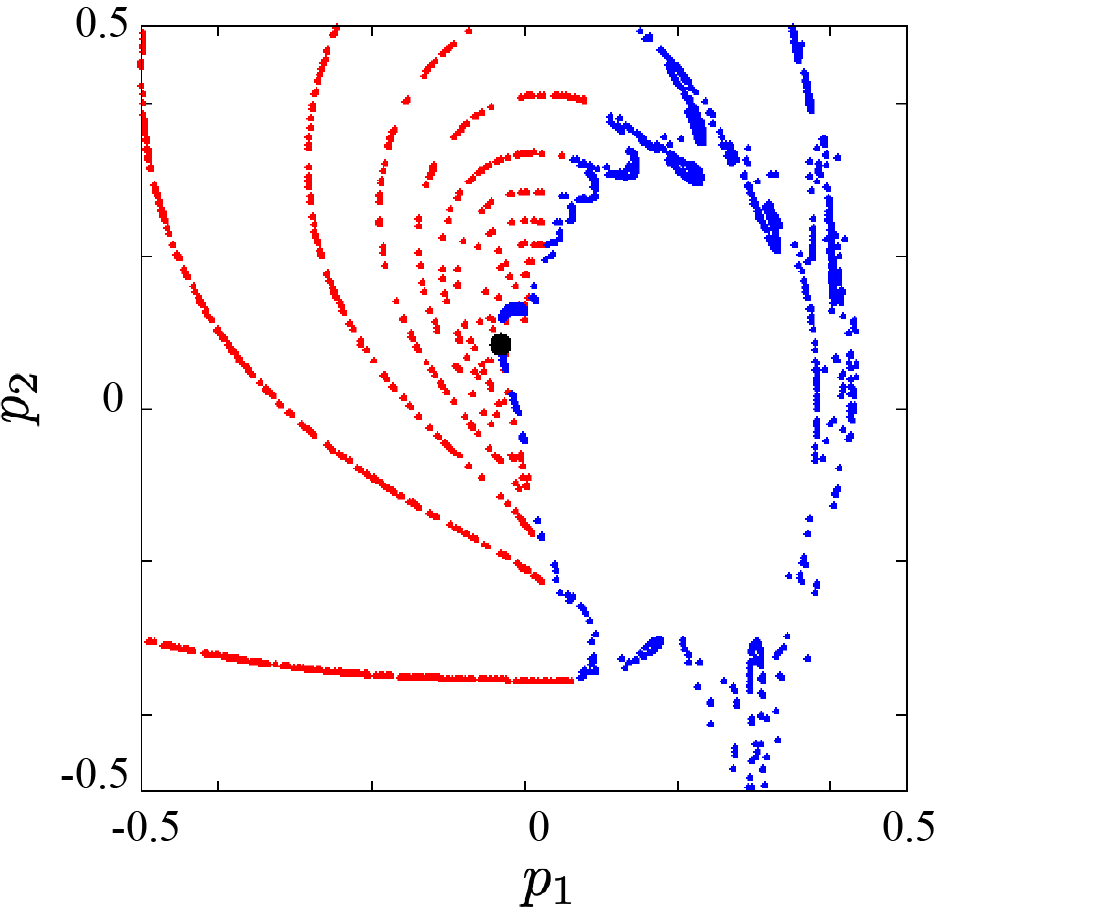}
      \caption{The contour plot of the initial conditions with the terminal time $t_f=50$, and with the terminal state $q_2(t_f)=0.0$.} 
      \label{fig:tpbvp2}
    \end{subfigure}
      \caption{%Enlarged terminal-time/state diagram around a solution of the fixed-terminal-time TPBVP: 
      (a) Enlargement of the terminal-time diagram in Fig. 6(a), and (b) enlargement of the terminal-state diagram in Fig. 7(a), around the initial adjoint condition satisfying \(q_1(t_f)=\pi\), \(q_2(t_f)=0\), and \(t_f=50\). 
      (c) The contour plot of the initial conditions with the terminal time $t_f=50 \pm 0.01$, and with the terminal state $q_2(t_f)=0.0 \pm 0.01$, are plotted by blue and red points, respectively. The black circle denotes the initial adjoint condition obtained by solving the TPBVP with the standard Newton method in all panels. 
      %The set of the same terminal time/state  forms the basins. 
      }
    \label{fig:tpbvp}
\end{figure}

For a prescribed fixed-terminal-time, fixed-terminal-state optimal control problem, the initial adjoint condition must simultaneously satisfy the specified terminal time and terminal state. Hence, in the $(p_1(0),p_2(0))$ plane, a solution of the TPBVP corresponds to an intersection of the level sets of $t_f$ and $q_2^*$. As an example, Fig.~\ref{fig:tpbvp} illustrates the solution with $q_1(t_f)=\pi,~q_2(t_f)=q_2^*=0.0,~ t_f=50$.
Figures~\ref{fig:tpbvp}(a) and \ref{fig:tpbvp}(b) show enlarged regions of the terminal time and terminal state diagrams, respectively, around the corresponding solution. The black dot denotes the initial adjoint condition obtained by solving the TPBVP with the standard Newton method, which corresponds to the
intersection of the level sets $t_f=50$ and $q_2^*=0.0$. The NHIMs at $q_1=\pm\frac{\pi}{2}$ act as escape channels for this slow transport, selecting the left or right exit toward the terminal state at $q_1=\pm\pi$. This dynamical behavior highlights the hypersensitivity of the optimal control problem: achieving the prescribed terminal time $t_f$ and the terminal state $q_2^*$ require fine tuning of the initial conditions.

%%%%%%%%%%%%%%%%%%%%%%%%%%%%%%%%%%%%%%%%%%%%%%%%%%%%%%%%%%%%%%%%%%%%%%%%%%%%%%%%
\section{CONCLUSIONS}

We show that a degenerate center acts as a turnpike, while normally hyperbolic invariant manifolds (NHIMs) serve as escape channels in the optimal control of the inverted pendulum. Spiral-like slow escape from the degenerate center arises from prolonged stagnation near the center before exiting via NHIMs. 

As a consequence of the finite terminal time, small perturbations in the initial adjoint variables lead to qualitatively distinct optimal trajectories, resulting in fractal structure of "exit basins". Such structure %are ubiquitous 
in optimal control problems pose severe challenges for trajectory optimization. Finite-time and finite-precision effects, together with hypersensitivity and the turnpike phenomenon, play a central role in limiting trajectory optimization. The restrictions with finite-time and finite-precision induces fractal exit basins and hypersensitivity, thereby linking the turnpike phenomenon to phase-space transports in optimal control. Details of the global state space geometry, the fractal dimensions of exit  basins, and the precise statistics of escape time from the turnpikes will be reported elsewhere.

Although the analysis is demonstrated on the inverted pendulum, the presented interpretation of hypersensitivity and turnpikes is not restricted to this example. The underlying mechanism, namely the interplay between slow dynamics near invariant sets and escape channels represented by NHIMs, is relevant to a broad class of optimal control problems. 
%From a computational viewpoint, the present analysis provides a geometric explanation for the ill-conditioning encountered in solving two-point boundary value problems. 
%The fractal structure of exit basins implies that small perturbations in the initial adjoint variables can lead to qualitatively different trajectories, making standard shooting methods highly sensitive. 
%This insight suggests that incorporating phase-space structure, such as invariant manifolds, may be essential for developing robust numerical algorithms.
%%%%%%%%%%%%%%%%%%%%%%%%%%%%%%%%%%%%%%%%%%%%%%%%%%%%%%%%%%%%%%%%%%%%%%%%%%%%%%%%
%\section{ACKNOWLEDGMENTS}
%
%The authors gratefully acknowledge the contribution of Organization and reviewers' comments. 

%%%%%%%%%%%%%%%%%%%%%%%%%%%%%%%%%%%%%%%%%%%%%%%%%%%%%%%%%%%%%%%%%%%%%%%%%%%%%%%%

\bibliographystyle{IEEEtran}
\bibliography{cdc}

\end{document}